\documentclass{paper}
\usepackage{mathrsfs}
\usepackage{graphics}
\usepackage{tikz}
\usepackage{pgfplots}
\usepackage{verbatim}
\usepackage{standalone}
\usepackage{dashrule}
\pgfplotsset{compat=1.9}

\begin{document}

%
%

%


%
%

%

%
\title{The critical layer during transition of the spiral Poiseuille flow in an annular gap}
\author{
 Venkatesa I. Vasanta Ram \\ Institut f\"ur Thermo- und Fluiddynamik \\ Ruhr Universit\"at Bochum, 44780 Bochum, Germany}

  
\maketitle

\keywords{flow transition, annular spiral flow, critical layer}
\begin{abstract}
%
The subject of this paper is on the propagation of disturbances causing transition in a fully developed spiral annular flow. The problem is approached through reformulation of the 
linearised equations of motion governing small-amplitude disturbances in a parameter space that differs from what is conventionally employed for this purpose. The alternative 
parameter space comprises a suitably defined Reynolds number formed with a {\it resultant characteristic velocity} that is a vectorial sum of the axial and azimuthal 
characteristic velocities of the problem, a {\it swirl parameter} that is the ratio of the azimuthal to the axial characteristic velocity, and the ratio of the annular gap to the 
mean diameter of the cylinders. In the limits of the swirl parameter assuming very small or very large values, the newly derived {\it generalised Orr-Sommerfeld and Squire} 
equations for disturbance propagation in the revised parameter space reduce to the corresponding equations that are known and well established for the limiting cases of disturbance 
propagation as the {\it swirl parameter} goes to zero or infinity respectively. Furthermore, they also lead naturally both to the criterion that determines from the equations 
themselves the analytical expression for the {\it location of the critical layer} in spiral annular flow, and to the equation governing the {\it flow dynamics in the critical layer}, 
thus revealing the effect of swirl on the critical layer.

\end{abstract}

%
\section{Introduction} 
The spiral flow in the annulus between concentric circular cylinders is an example 
of a flow that is susceptible to transition through combined action of several 
physical mechanisms. In the particular case when the helical pattern of the basic flow streamlines that characterises a spiral flow results from an imposed axial pressure gradient 
in conjunction with rotation of the inner cylinder alone, there are two well-known distinct physical mechanisms of disturbance propagation that can trigger transition. These are: 
1) the {\it Tollmien-Schlichting mechanism} and, 2) the {\it Taylor mechanism}. The features of these mechanisms acting on their own come convincingly to light when transition of 
the spiral flow in question in its limiting cases are contrasted with each other. The limiting cases are when the pitch of the helix, measured through the angle the helix makes 
with the common axis of the cylinders, assumes values of either $ 0 $ or $\frac{\pi}{2}$. Differences between the {\it Tollmien-Schlichting mechanism} and the {\it Taylor mechanism} 
are indeed striking and readily visible in flow visualization studies, see eg. \cite{albumvandyke}. In theory the differences manifest themselves most convincingly when dynamical 
characteristics of transition inducing disturbances in the limiting cases of the spiral flow are contrasted with each other. An outstanding difference between the two limiting cases 
is the appearance of a {\it critical layer} within the flow when the {\it Tollmien-Schlichting mechanism} alone is responsible for transition, which is realised when the pitch of 
the helix is $0$. In contrast, no {\it critical layer} appears when the {\it Taylor mechanism} alone causes transition as it is when the pitch is $\frac{\pi}{2}$. From physical 
reasoning based on affinity of disturbance propagation characteristics, one would expect that a {\it critical layer} might arise in the spiral flow undergoing transition when the 
pitch of the helix of its basic flow streamline is close to $0$. It would vanish on the pitch increasing from $0$, to disappear entirely when the pitch reaches $\frac{\pi}{2}$. It 
is then of scientific interest to examine when and/or under what conditions a {\it critical layer} may arise or disappear during transition of the spiral flow. The aim of the 
present work is to derive these conditions from first principles, {\it i.e.} from the linearised equations governing propagation of small-amplitide disturbances to the spiral flow.    
   
The significance of the {\it critical layer} in transition is related to the generation of {\it Reynolds stresses}, see eg. \cite{stuart}, \cite{cclin}, \cite{drazinreid},
\cite{maslowe1}, \cite{maslowe2}, \cite{schmidhenningson}, \cite{ramagovindarajan}, \cite{criminalejacksonjoslin}. Besides the differences in the mechanisms referred to earlier 
some further pronounced differences may be outlined. These are as follows. In the {\it Tollmien-Schlichting mechanism} transition is induced by travelling waves of the 
disturbance. These are effective even in a basic flow with straight and parallel streamlines. Curvature of the basic flow streamlines is therefore not essential for onset of 
transition through this mechanism. In contrast, in the {\it Taylor mechanism} transition is not possible in the absence of curvature of the basic flow streamlines. When the 
{\it critical layer} appears, it is located at a position in the flow where flow velocity and the phase velocity of the neutrally stable travelling disturbance wave are equal to 
each other. Its location coincides with a singularity occuring in the inviscid form of the set of linearized equations for disturbance propagation. The singularity is most readily 
evident when the set of linearised equations for disturbances are cast in a form of the {\it Orr-Sommerfeld equation}. It is then recognisable through the coefficient of the second 
derivative of the wall-normal component of the disturbance velocity changing its sign.   

\subsection*{Outline of the present paper}
In Section 2 the present state of understanding of the {\it critical layer} during transition of spiral Poiseuille flow is subject to a brief critical review. The purpose of this 
Section is to prepare ground for the reformulation of the disturbance propagatioon problem in a spiral Poiseulle flow undertaken in the present work. 
It will be seen in the course of this work that a {\it reformulation} of disturbance propagation is essential for proper imbedding of the phenomenon in the flow of current interest 
in the extensively studied past work of disturbance propagation in its two limiting cases. 
In Section 3 the reformulation of the problem is introduced, which is essentially the derivation of the 
{\it Generalised Orr-Sommerfeld and Squire equations}. In Section 4 the relations of the {\it Generalised Orr-Sommerfeld and Squire equations} to the limiting cases of flows
with mild and strong swirl are illuminated. The equations for the case of mild swirl bring out the outstanding characteristics of the {\it critical layer in spiral flow} in an 
analytical form. In Section 5, the meaning of the analytical results derived are discussed and summarised. 

\section{A brief critical review of the present state of understanding of the {\it critical layer} during transition of the spiral Poiseuille flow}
When both the {\it Tollmien-Schlichting mechanism} and the {\it Taylor mechanism} are acting simultaneously, as in the {\it annular spiral flow}, questions regarding the dominance 
of one over the other in a certain region of the parameter space arise. Forms of manifestation of this dominance are, as expected, of fundamental scientific interest. Physical reasoning would lead one 
to expect that the {\it Tollmien-Schlichting mechanism} would dominate over the other when the pitch of the helix is close to $0$, whereas it would be the {\it Taylor mechanism} 
when the pitch is close to $\frac{\pi}{2}$. This expectation would in turn lead to the possibility of a {\it critical layer} arising in the spiral flow 
when its helical angle lies close to $0$. Besides the mechanism as such, knowledge of the character of the dominant transition inducing disturbance, in particular whether it is 
toroidal or helical in nature, would shed light on the changes that might be expected in the observable flow 
pattern as the flow is taken through the parameter space, thus forming a major step towards understanding of the complexity of transition in this apparently simple flow.  
It is therefore only but natural that a large number of research papers are devoted to studies of disturbance propagation and transition in spiral flows. A 
cursory search through published literature shows them to have been appearing from time to time, starting from the early days of transition research, see eg. the classical book of 
Chandrasekhar \cite{chandrasekhar}, 
and continuing right into the present day. An insight into the highly ramified nature of the problems arising in this and related flows is provided by the 
papers of Hasoon and Martin \cite{hasoonmartin}, DiPrima and Pridor \cite{diprimapridor}, Takeuchi and Jankowski \cite{takeuchijankowski}, Ng and Turner \cite{ngturner}, 
papers authored or co-authored by Cotrell and Pearlstein \cite{cotrellpearlstein}, by Meseguer and Marques \cite{meseguermarques1, meseguermarques2, meseguermarques3}, by Avila, 
Meseguer and Marques \cite{avilameseguermarques}, \cite{meseguermellibovskyavilamarques}, \cite{marquesmeseguerlopezetal}, by Leclerq, Pier and Scott \cite{leclercqpierscott}, 
Deguchi and Altmeyer \cite{deguchialtmeyer}, \cite{altmeyer}, and by the authors of further references cited in these papers. 

An examination of the equations for the study of disturbance propagation which are employed in the cited references shows that the approaches taken by the authors may be grouped 
under two broad headings depending upon the manner in which the pressure disturbance is treated. In one, the pressure disturbance is treated explicitly as an unknown, which leaves 
four unknowns to be determined from the four equations that are available, {\it viz.} the momentum equations in the three directions and the continuity equation. In the other, the 
pressure disturbance does not appear as an unknown, and there are only three unknowns which are the components of the velocity disturbance. The task then faced is to formulate an 
appropriate set of three equations for determining the three components of the velocity disturbance, without loss of substance of the subject of investigation. A study of 
published literature shows that this task is carried out through procedures tailored to the basic flow. For the basic flow with straight and parallel streamlines it is the 
procedure for derivation of the {\it Orr-Sommerfeld equation}. For the basic flow with concentric circular streamlines it is the one worked out by Taylor for the 
{\it Taylor stability problem}, see  Taylor \cite{taylorscientificpapers}. Both the procedures are described in well known published literature, eg. Stuart \cite{stuart}, 
Schmid and Henningson \cite{schmidhenningson}, Criminale, Jackson and Joslin \cite{criminalejacksonjoslin}, Chossat and Iooss \cite{chossatiooss}, Meyer-Spasche \cite{meyerspasche}. The relation between the two 
apparently diverse derivation procedures for elimination of the pressure disturbance in these two basic 
flows does not seem to have been a subject for discussion in published literature. A pressure disturbance elimination procedure for the basic flow of present interest which is 
characterized by helically wound streamlines, that is worked out along lines analogous to those in the other kinds of basic flow referred to, is also, to the knowledge of the 
authors, not known from published literature. One of the objectives of the present work is to present such a derivation 
procedure for the {\it spiral Poiseuille flow} that bridges the procedures for elimination of the pressure disturbance as an unknown in basic flows with 
{\it straight and parallel streamlines} on one hand and {\it concentrically circular streamlines} on the other.

At this juncture it is in order to draw attention to some salient differences between the equations for the disturbance formulated in the two approaches just mentioned, when they 
are applied to the two limiting cases considered. In the approach with the pressure disturbance retained as an unknown, derivatives of the velocity disturbance appear up to the second order, 
whereas for the pressure disturbance they are present only to the first order. In the approach with the pressure disturbance eliminated, fourth order derivatives of the 
velocity disturbance appear for both the limiting cases, however with a subtle structural difference between the two. For the limiting case with the basic flow described by 
straight and parallel streamlines, the coefficient of the term with a second-order derivative of the velocity disturbance may pass through a zero, whereas there is no such 
zero-crossing in the equation for the velocity disturbance in the case with concentrically circular streamlines in the basic flow. It is now well known that the zero-crossing in the coefficient of the second derivative is decisive for the 
occurence of a {\it critical layer}, {\it viz. Tollmien's critical layer}, its location being given by this zero-crossing as an asymptotic approximation. As a matter of fact, it is 
the zero-crossing of the coefficient of the second derivative of the velocity disturbance that renders the occurence of the critical layer transparent from the governing equations 
themselves. As against this, no such critical layer arises in the corresponding equations for the basic flow with concentric circular streamlines. Neither is the critical layer 
discernible from the governing equations themselves if the pressure disturbance is retained as an unknown in their formulation.  

In {\it spiral Poiseuille flow}, the {\it strength} of swirl may be characterised through the ratio of the {\it two characteristic velocities} of the basic flow, which are the 
{\it axial} and {\it azimuthal characteristic velocities}, denoted $U_{refx}$ and $U_{ref \varphi}$ respectively. When the two mechanisms are acting simultaneously, as in the flow 
in question, intuition would suggest that the disturbance propagation mechanism of the transition inducing disturbance would lie closer to either the {\it Tollmien-Schlichting} or 
the {\it Taylor} mechanism, depending upon the ratio of the two characteristic velocities. One would then expect that, for an arbitrarily given value of this ratio, which may lie 
between between zero and infinity, 
the set of linearised equations themselves that describe the velocity disturbances in the swirling 
basic flow in the annulus would lie closer to the corresponding set of equations for the velocity disturbances when the ratio of the two characteristic 
velocities attains values of zero or infinity, as the case may be. 
These are the Orr-Sommerfeld and Squire equations for the plane channel flow, possibly corrected for the effects 
of transverse curvature, when the azimuthal characteristic velocity is zero, and, Taylor's equations for disturbance propagation in the annular gap between concentric cylinders 
when the axial characteristic velocity is zero. 
A close surveillance of the published work in this area, however, shows that, when the approach with elimination of the pressure 
disturbance is followed, the equations employed in the cited literature to describe disturbance propagation in swirling flows do not degenerate to the Orr-Sommerfeld and Squire 
equations in the absence of swirl in the basic flow. 

\section{The proposed alternative formulation}
The basic flow of our problem is the fully developed spiral flow in an annulus. The proposed alternative approach views the phenomenon of disturbance propagation in this 
flow in terms of a set of revised parameters which are going to be shortly introduced. Conventionally, the axial and azimuthal velocity components in the 
basic flow of our problem, $V_{Gx}$ and $V_{G \varphi}$, are written as products of the two characteristic velocities in the two directions, {\it viz.} $U_{refx}$ and $U_{ref\varphi}$ 
respectively, with the shape functions in that order, $U_x (r; \epsilon_R)$ and $U_{\varphi} (r; \epsilon_R)$. Here $r$ is the radius, and $\epsilon_R = \frac{R_o - R_i}{R_o + R_i}$ is a 
measure of the transverse curvature with $R_o$ and $R_i$ denoting the 
radii of the outer and the inner cylindrical surfaces of the annulus. In the particular case of the spiral flow in question, in which the flow is maintained in the 
wall-bounded annulus between the concentric cylinders by an axial pressure gradient, $\frac{dP_G}{dx}$, together with the inner cylinder rotating with 
an angular velocity, $\Omega_i$, a natural choice for the characteristic velocities would be: $U_{refx}= - \frac{H^2}{2 \mu} \frac{dP_G}{dx}$ and $U_{ref\varphi} = R_i \Omega_i$. 
Here, $2H = R_o - R_i$ denotes the annular gap between the cylinders, and $\mu$ the dynamic viscosity. We may note that the profiles of the axial and the azimuthal velocity 
components of the basic flow, $U_{Gx}$  and $U_{G \varphi}$, depend upon the transverse curvature parameter $\epsilon_R$. They may also depend upon further parameters entering the 
problem, which might be the case when e.g. a cylinder wall is prescribed to move with a certain axial translational velocity. 

The revised parameters for our problem are the set constituting a characteristic velocity, $U_{ref}$, defined through $U_{ref}^2 = U_{refx}^2 + U_{ref \varphi}^2$, and a swirl 
parameter, $S$, defined through $S = \frac{U_{ref \varphi}}{U_{refx}}$. The velocity components of the basic flow are then given in terms of these revised parameters 
through the following expressions: 
\begin{eqnarray}
V_{Gx} = U_{ref} (1+S^2)^{- \frac{1}{2}} U_{Gx}(r; \epsilon_R); ~   ~   ~   V_{G \varphi} = U_{ref} S (1+S^2)^{- \frac{1}{2}} U_{G \varphi}(r; \epsilon_R).  
\label{basicflowrevisedparam}
\end{eqnarray} 
The starting point for our work is the set of equations of motion in cylindrical co-ordinates which may be found in standard text books on fluid dynamics, ({\it vide} e.g. 
\cite{schlichting}, \cite{batchelor}). Non-dimensionalisation of these equations with $U_{ref}$ and $H$ as reference quantities, setting the velocity and the pressure fields as sums of 
the basic flow and a disturbance, followed by linearisation for disturbances lead to equations for small-amplitude disturbances to the velocity and pressure fields of the basic flow. 
From these linearised equations for disturbances, the pressure disturbance may be eliminated through two different procedures which are described for the basic flow with straight 
and parallel streamlines in standard works on hydrodynamic stability {\it vide} e.g. \cite{schmidhenningson}. We refer to these procedures as the {\it Orr-Sommerfeld} and 
{\it Squire} procedures respectively. Carrying out these procedures in the cylindrical co-ordinate system adopted for the present problem leads in that order to the 
{\it generalised  Orr-Sommerfeld} and {\it Squire equations}. They are, together with the {\it Continuity equation}, the governing equations for our problem. The unknowns in this 
set of equations are only the three components of the velocity disturbance, since the disturbance to the pressure does not appear any longer as an unknown. Profiles of the axial 
and azimuthal components of the basic flow, $U_{Gx}(r)$ and $U_{G \varphi}(r)$, and their first and second derivatives occur as coefficients in the governing equations. These 
equations are given to the order $O(r^{-2})$ in the {\bf Appendix}. For the purposes of the present work it is convenient to write the set in a compact matrix form as follows: 
\begin{equation}
              D \mathbf u = \mathbf 0. 
\label{generalisedossqcont}
\end{equation}
We refer to (\ref{generalisedossqcont}) as the set of {\it generalised Orr-Sommerfeld, Squire and Continuity equations}.
Here the linear differential operator, $ D$, depends upon the Reynolds number $Re$ based on the characteristic velocity $ U_{ref}$ and the semi-gap width $ H $ as reference quantities, 
on the swirl parameter $S = \frac{U_{ref \varphi}}{U_{ref x}}$, and also upon the ratio of the annular gap-width to the mean cylinder radius, $\epsilon_R$. 
The differential operator $ D$ itself may be 
written in a matrix form as follows: 
\begin{eqnarray}
 D  = \left(
\begin{array}{ccc}
  D_{OSr}   &   D_{OS \varphi}   &   D_{OSx}\\
  D_{Sqr}   &   D_{Sq \varphi}   &   D_{Sqx}\\
  D_{Cor}   &   D_{Co \varphi}   &   D_{Cox}
\end{array}
\right),   
\label{breakdownofossqco}
\end{eqnarray} 
where the subscripts $OS$, $Sq$ and $Co$ indicate in that order the matrix partial differential operators arising from the {\it generalised Orr-Sommerfeld}, {\it Squire} and Continuity 
equations, and the subscripts $r, \varphi$ and $x$ stand for partial derivatives operating on $u_r, u_{\varphi}$ and $u_x$ respectively. Expressions for the operators in 
(\ref{breakdownofossqco}) follow from a comparison of (\ref{breakdownofossqco}) with (\ref{generalisedorrsommerfeld}), (\ref{generalisedsquire}) and the continuity equation given in the {\bf Appendix}. We may draw the reader's 
attention at this juncture to a salient difference between the {\it generalised Orr-Sommerfeld equation}, and {\it Orr-Sommerfeld equation} for the conventional problem. It is that, 
in contrast to the conventional case in which disturbance propagation in a basic flow with wall-parallel streamlines is described, in the {\it generalised Orr-Sommerfeld equation}, the 
wall-normal component of the velocity disturbance $u_r$ does not stand all by itself. The transverse curvature inherent in the geometry causes $u_r$ to occur coupled with $u_{\varphi}$. 
\section{The relation between the {\it generalised Orr-Sommerfeld and Squire equations} and the equations for the classical cases.}
It will be shown in this section that the {\it generalised Orr-Sommerfeld and Squire equations}, (\ref{generalisedossqcont}), which are written in terms of the revised parameters, 
degenerate into the known classical cases when the swirl parameter, $S$, approaches the limits of $0$ or $\infty$. These are the  well-established equations 
describing the transition mechanisms of {\it Tollmien-Schlichting} and {\it Taylor} respectively. Besides 
establishing these relationships, the equations for the limit $S \rightarrow 0$ also show in analytical form the nature of the dependence of the location of the critical layer 
on the parameter $S$. For illustrative purposes we will demonstrate this degeneration process for the example of the fully developed spiral flow in an annulus whose values of 
the ratio of the gap width to the mean radius is small, and in conjunction with rotation of the inner cylinder alone. To this end, it is meaningful to transform the variable from the radius $r$ to $y$, where $y$ 
is the radial co-ordinate measured from the mean radius $\frac{R_o + R_i}{2}$, non-dimensionalised with $H$, and convert the partial differential equation (\ref{generalisedossqcont}) 
into an ordinary differential equation. The conversion may be carried out through substitution of the standard modal ansatz 
\begin{equation}
\mathbf u = \mathbf A(y) \exp \{\imath (\lambda_x x + n_{\varphi} \varphi) - \omega t \} + c.c.,
\label{modalansatzforu}
\end{equation} 
in (\ref{generalisedossqcont}). In (\ref{modalansatzforu}), $\bf A$ is a column vector whose components are, in a self-explanatory notation, $(A_r, A_{\varphi}, A_x)^T$. Using  
(\ref{modalansatzforu}) in conjunction with asymptotic expansions for $U_{Gx}$ and $U_{G \varphi}$ for small values of the geometrical parameter 
$\epsilon_R = \frac{R_o - R_i}{R_o + R_i} = \frac{2 H}{R_o + R_i} \rightarrow ~ 0$, which may be verified to be $ \left(U_{Gx} \right)_{\epsilon_R \rightarrow 0} \simeq \left((1-y^2) - \epsilon_R \frac{y(1-y^2)}{3} + O(\epsilon_R^2) \right)  $ 
and  
$ \left(U_{G \varphi}\right)_{\epsilon_R \rightarrow 0} \simeq \left(\frac{(1-y)}{2} - \epsilon_R \frac{(1-y^2)}{2} + O(\epsilon_R^2) \right)$, 
casts (\ref{generalisedossqcont}) into the set of ordinary differential equations for $(A_r, A_{\varphi}, A_x)^T$.  
In order to facilitate comparison of the different derivatives of $\mathbf A$ in the {\it generalised Orr-Sommerfeld, Squire and Continuity equations} with corresponding terms in 
plane channel flow it is meaningful to rewrite the converted set of ordinary differential equations ordered according to the coefficients of the vectorial velocity disturbance and 
its derivatives according to the following scheme: 
\begin{equation}
   \mathbf K_4 \frac{d^4 \bf A}{dy^4} + \mathbf K_3 \frac{d^3 \bf A}{dy^3} + \mathbf K_2 \frac{d^2 \bf A}{dy^2}+ \mathbf K_1 \frac{d \bf A}{dy} + \mathbf K_0 \bf A = 0.  
\label{reorderedgeneralisedossqcont}
\end{equation}
In (\ref{reorderedgeneralisedossqcont}), $\mathbf K_4, . . \mathbf K_0 $ are matrices that are derivable from the matrix differential operator $ D$, together with the 
ansatz (\ref{modalansatzforu}). They may themselves be grouped as follows:
\begin{eqnarray}
\mathbf K_k  = \left(
\begin{array}{ccc}
  K_{kOSr}   &   K_{kOS \varphi}   &  K_{OSx}\\
  K_{kSqr}   &   K_{Sq \varphi}    &  K_{Sqx}\\
  K_{Cor}    &   K_{Co \varphi}    &  K_{Cox}
\end{array}
\right),   
\label{defnmathbfk}
\end{eqnarray} 
where the subscript $k$ stands for $k = 0, 1, 2, 3$ or $4$. Many elements of the coefficient matrices $\mathbf K_0, . . . \mathbf K_4 $ are zero. 

From the {\it Orr-Sommerfeld} row of the equations (\ref{reorderedgeneralisedossqcont}, \ref{defnmathbfk}) we may obtain, on division through $(\imath \omega)$, 
the ordinary {\it generalised Orr-Sommerfeld} equation that can be written in a form that clearly brings out the similarities and differences between the classical 
case of the plane channel flow and the annular spiral flow presently under consideration. We get, including terms to the order $O(\epsilon_R)$: 
\begin{eqnarray}
 \frac{1}{\imath \omega} \frac{1}{Re} \left(\frac{d^4 A_r}{dy^4} \right) + \left(1- \frac{\lambda_x}{\omega} \frac{(1-y^2)}{\sqrt{(1+S^2)}} - \imath \frac{2}{Re}\frac{\lambda_x^2}{\omega} \right) \frac{d^2 A_r}{dy^2}    \nonumber   \\ 
  - \left(1 + \frac{\lambda_x}{\omega} \frac{2}{\sqrt{(1+S^2)}} - \frac{\lambda_x^3}{\omega} \frac{(1-y^2)}{\sqrt{(1+S^2)}} \right) A_r    \nonumber   \\ 
- \epsilon_R \left[\frac{\imath}{ \omega} \frac{2}{Re} \frac{d^3A_r}{dy^3} + \left( \frac{\lambda_x}{\omega} \frac{1}{\sqrt{(1+S^2)}} \frac{y(1-y^2)}{3} + \frac{n_{\varphi}}{\omega} \frac{S}{\sqrt{(1+S^2)}} \frac{(1-y)}{2} \right) \frac{d^2 A_r}{dy^2}  \right]  \nonumber \\    
 + \epsilon_R \left[1 - \frac{\lambda_x}{\omega} \frac{(1-y^2)}{\sqrt{(1+S^2)}} + \imath \frac{2 \lambda_x^2}{\omega Re}  \right] \frac{dA_r}{dy}    \nonumber  \\  
 + \epsilon_R \left[\frac{\lambda_x}{\omega} \frac{1}{\sqrt{(1+S^2)}}\left(4y + \frac{y(1-y^2) \lambda_x^2}{3} \right) +  \frac{n_{\varphi}}{\omega} \frac{S}{\sqrt{(1+S^2)}} \frac{(1-y)}{2} \lambda_x^2 \right] A_r  = 0. 
\label{oseqnforar}
\end{eqnarray}

%
%
%
%

A cursory examination of the above ordinary differential equation shows that for $\epsilon_R = 0$, (\ref{oseqnforar}) reduces to the Orr-Sommerfeld equation for plane channel flow 
that is known from literature, eg. \cite{schlichting}, \cite{stuart}, 
\cite{schmidhenningson}, \cite{criminalejacksonjoslin}. For $\epsilon_R \ne 0 $ corrections to the classical Orr-Sommerfeld equation arise and a surveillance of the correction terms shows that among these, two different 
forms of dependence on the swirl parameter, $S$, are distinguishable. In one of these $\frac{1}{\sqrt{(1+S^2)}}$ appears as a coefficient, and this contribution is ascribable to the 
transverse curvature that is inherent in the geometry considered. The other exhibits $\frac{S}{\sqrt{(1+S^2)}}$ as a coefficient, and this describes the effect of swirl. Terms 
belonging to the former remain small over the entire range of values of $S$ from $0$ to $\infty$, whereas those belonging to the latter need not remain small when 
$S \rightarrow \infty$. We will refer to these two as cases of small and large swirl respectively, and examine them more closely.   

\subsection{The case of small swirl and the critical layer}
When the corrections to the equations are small, which is the case as long as the product $\epsilon_R S n_{\varphi}$ remains small, we may expect the departure of the solutions 
from the case $S = 0$ also to remain small, as long as there are no changes in the other parameters, $Re, \epsilon_R$ and in the mode $n_{\varphi}$. This is tantamount to the expectation 
that the neutral stability curves in the $(Re,\lambda_x)$-plane in the $(Re, S, \epsilon_R; \lambda_x, n_{\varphi})$-space bear similarity to each other, a feature that would 
facilitate computational determination of the neutrally stable surface and the critical parameters in the multi-parameter space of $Re, S$ and $\epsilon_R$. We may note that the 
equation (\ref{oseqnforar}) that is necessary to be solved to reach this objective, is an eigenvalue problem in which $A_r$ is the only unknown, without any coupling with 
$A_{\varphi}$ or $A_x$. Numerical methods for solving the Orr-Sommerfeld equation in the more classical problems of hydrodynamic stability that have been developed, tested, and 
are available in published literature, see eg. \cite{schmidhenningson}, \cite{criminalejacksonjoslin}, may then be expected to be applicable for solving (\ref{oseqnforar}). 
\subsubsection{The critical layer, its location and scaling properties}
The {\it Orr-Sommerfeld equation} for the case of mild swirl, (\ref{oseqnforar}), on comparison with the corresponding equation for the case of the basic flow with straight and 
parallel streamlines, see eg. \cite{criminalejacksonjoslin}, exhibits a singularity of the same nature as in the classical case. It arises for neutrally stable disturbances in the 
inviscid counterpart of the problem presently under study when the Reynolds number is large. This observation leads to the expression for the location of the {\it critical layer} 
in the flow in question. Analogous to the classical case, this singularity is removable in the complete problem through introduction of a thin {\it critical layer} of appropriate 
thickness in which viscous effects are retained. In the classical case of a basic flow with straight and parallel streamlines this is the well-known {\it Tollmien's critical layer}, 
and its thickness is $O(Re^{-\frac{1}{3}})$. Application of the same procedure yields the scaling behaviour of the {\it critical layer} for the flow in question. 

\subsubsection*{Location of the critical layer}
Inspection of (\ref{oseqnforar}) shows that the coefficient of the second derivative of $A_r$ goes through zero for neutrally stable 
disturbances when the Reynolds number $Re$ is not small, as it does in the classical case. The criterion for the {\it location of the critical layer}, $y = y_c$, may therefore be 
written as follows: 
\begin{equation}
 1 - \frac{\lambda_x}{\omega} \frac{(1-y_c^2)}{\sqrt{(1+S^2)}} 
- \epsilon_R  \left( \frac{\lambda_x}{\omega} \frac{1}{\sqrt{(1+S^2)}} \frac{y_c(1-y_c^2)}{3} + \frac{n_{\varphi}}{\omega} \frac{S}{\sqrt{(1+S^2)}} \frac{(1-y_c)}{2} \right) = 0.    
\label{critlayerloc}
\end{equation}
For $\epsilon_R = 0$ the above expression is seen to be identical with the known result, i. e. the critical layer is located at a wall distance where the basic flow velocity and 
the phase speed of neutrally stable disturbances equal each other, see eg. \cite{criminalejacksonjoslin}. For $\epsilon_R \ne 0 $ the location 
of the critical layer depends, besides on $S$, also upon the mode of the disturbance $n_{\varphi}$. It is only for toroidal modes, for which $n_{\varphi} = 0$, that (\ref{critlayerloc}) 
fulfills the condition of the local flow velocity and the phase velocity of the wave equalling each other, cf. expression for $U_{Gx}$ in the text following 
(\ref{modalansatzforu}). For the 
modes $n_{\varphi} \ne 0$ the terms subtracted from $1$ in the expression (\ref{critlayerloc}) may be interpreted as the {\it inner product} of the {\it basic flow velocity vector},  
$(U_{Gx}, U_{G \varphi})$, and the {\it slowness vector} of the wave of the particular mode in question, $\frac{(\lambda_x, n_{\varphi})}{\omega}$, a concept introduced by 
Whitham in \cite{whitham}.

\subsubsection*{The order of magnitude of the critical layer thickness}
An estimate of the order of magnitude of the critical layer thickness is obtainable by application of the same procedure as described in literature, eg. \cite{criminalejacksonjoslin}), to the 
present problem. To this end we introduce in the critical layer the independent variable $\eta_c$ defined through $\eta_c = (y-y_c) Re^m_c$, write (\ref{oseqnforar}) in the 
neighbourhood of the critical layer and determine $m_c$ such that the coefficients of $\frac{d^4 A_r}{dy^4}$ and of $\frac{d^2 A_r}{dy^2}$ in (\ref{oseqnforar}) can stand in balance 
with each other. Carrying out this 
procedure for the present problem shows that a distinction has to be made between the {\bf modes of the disturbance}, i.e. between disturbances that are {\it toroidal} for which 
$n_{\varphi}= 0$, and those that are {\it helical} for which $n_{\varphi} \ne 0$. For {\it toroidal} disturbances one gets the same result as in the conventional case which is  
$m_c = -\frac{1}{3}$.  This result for $m_c$ remains applicable to {\it helical disturbances}, $n_{\varphi} \ne 0$, too, only as long as the product $\epsilon_R S n_{\varphi}$ 
remains numerically small. The equation governing $A_r$ in the {\it critical layer} is then as follows: 

\begin{eqnarray}
 \frac{1}{\imath \omega} \left(\frac{d^4 A_r}{d\eta_c^4} \right) + \left(1+ \frac{\lambda_x}{\omega} \frac{1}{\sqrt{(1+S^2)}}(2y_c \eta_c)  \right) \frac{d^2 A_r}{d\eta_c^2}    \nonumber   \\ 
%
+ \epsilon_R \left[ \left( \frac{\lambda_x}{\omega} \frac{1}{\sqrt{(1+S^2)}} \frac{(2y_c \eta_c)}{3} + \frac{n_{\varphi}}{\omega} \frac{S}{\sqrt{(1+S^2)}} \frac{\eta_c}{2} \right) \frac{d^2 A_r}{d\eta_c^2}  \right]      
%
%
\label{oseqnforarincritlayer}
\end{eqnarray}

%
%
%

%
%

\subsection{The case of large swirl}
It was seen in the previous subsection that the {\it generalised Orr-Sommerfeld and Squire equations}, (\ref{generalisedossqcont}) or (\ref{generalisedorrsommerfeld}, \ref{generalisedsquire}), 
degenerate to their classical counterparts in plane channel flow, with corrections due to transverse curvature, as the {\it swirl parameter}, $S$, goes to $0$. In contrast, 
in the other limit $S \rightarrow \infty $, a close examination of the same set of equations would show that the equations ( \ref{generalisedorrsommerfeld}, \ref{generalisedsquire}) 
then do not degenerate to those derived by Taylor for his stability problem. The noteworthy aspect of this failure is that proceeding to the limit $S \rightarrow \infty$ in a 
straightforward manner does not result in an eigenvalue problem with a mutual coupling of $A_r$ and $A_{\varphi}$ deciding its stability, as it is the case in Taylor's treatment of this 
stability problem. However, this shortcoming may be overcome through rewriting the equations (\ref{generalisedorrsommerfeld}, \ref{generalisedsquire}) in {\it Taylor variables}, 
$(\hat t, \hat y, \hat \varphi, \hat u_r, \hat u_{\varphi}, \hat u_x)$, defined through 
\begin{equation}
\hat t = \frac{t}{Re}; \hat y = y; \hat \varphi = \varphi;  \hat x = x;   \hat u_r = u_r \frac{Re}{\epsilon_R}; \hat u_{\varphi} = u_{\varphi}; \hat u_x = u_x \frac{Re}{\epsilon_R} , 
\label{defntaylorvariables}
\end{equation}
before proceeding to the limit $S \rightarrow \infty$. One then gets the {\it generalised Orr-Sommerfeld, Squire and Continuity equations in Taylor variables} which we write as 
\begin{equation}
             \hat  D \mathbf {\hat u} = \mathbf 0,  
\label{generalisedossqcontintaylorvariables}
\end{equation}
where the expressions for the elements of the operator $ \hat  D$ in (\ref{generalisedossqcontintaylorvariables}) follow from a comparison of this equation with its detailed breakdown 
given below. 
\\
{\bf The generalised Orr-Sommerfeld equation for large swirl}
\begin{eqnarray}
\frac{\partial}{\partial \hat t} \left[ \left(-\frac{\partial^2 \hat u_r}{\partial \hat y^2} - \frac{\partial^2 \hat u_r}{\partial \hat x^2} \right) - \epsilon_R \frac{\partial \hat u_r}{\partial \hat y} + Re \left(\frac{\partial^2 \hat u_{\varphi}}{\partial \hat y \partial \hat \varphi} - \frac{\partial \hat u_{\varphi}}{\partial \hat \varphi}\right) \right] \nonumber  \\ 
+ \frac{U_{Gx}}{\sqrt{(1+S^2)}} Re \left[ \left(- \frac{\partial^3 \hat u_r}{\partial \hat y^2 \partial \hat x} - \frac{\partial^3 \hat u_r}{\partial \hat x^3} \right) + \epsilon_R \left( \frac{\partial ^3 \hat u_r}{\partial \hat y^2 \partial \hat \varphi} - \frac{\partial ^2 \hat u_r}{\partial \hat y \partial \hat x}\right) + Re \left( \frac{\partial ^2 \hat u_{\varphi}}{\partial \hat y \partial \hat \varphi} \right) \right]  \nonumber  \\ 
+ \frac{1}{\sqrt{(1+S^2)}} \frac{d U_{Gx}}{dy} \left[ \epsilon_R Re \frac{\partial \hat u_r}{\partial \hat x} + Re^2 \frac{\partial ^2 \hat u_{\varphi}}{\partial \hat \varphi \partial \hat x} \right] + \frac{1}{\sqrt{(1+S^2)}} \frac{d^2 U_{Gx}}{dy^2} Re \frac{\partial \hat u_r}{\partial \hat x}   \nonumber   \\ 
%
+ \frac{S }{\sqrt{(1+S^2)}} U_{G \varphi} \left[\epsilon_R Re \left(- \frac{\partial ^3 \hat u_r}{\partial \hat y^2 \partial \hat \varphi} - \frac{\partial ^3 \hat u_r}{\partial \hat x^2 \partial \varphi} \right) + 2 Re^2 \frac{\partial ^2 \hat u_{\varphi}}{\partial \hat x^2}  \right]   \nonumber   \\ 
+ \frac{S \epsilon_R}{\sqrt{(1+S^2)}} \frac{d U_{G \varphi}}{dy} Re \left[ \left( \frac{\partial ^2 \hat u_r}{\partial \hat y \partial \hat \varphi} + \frac{\partial ^2 \hat u_x}{\partial \hat x \partial \hat \varphi} \right) \right] + \frac{S \epsilon_R}{\sqrt{(1+S^2)}} \frac{d^2 U_{G \varphi}}{d y^2} \left[Re \frac{\partial \hat u_r}{\partial \hat \varphi} \right]  \nonumber   \\  
+ \left[ \left( \frac{\partial^4 \hat u_r }{\partial \hat x^4} + \frac{\partial^4 \hat u_r}{\partial \hat y^4} + 2 \frac{\partial^4 \hat u_r}{\partial \hat x^2 \partial \hat y^2} \right) + \epsilon_R Re \cdot 2 \frac{\partial^3 \hat u_r}{\partial \hat y \partial \hat x^2}  + Re \cdot 2 \frac{\partial ^3 \hat u_{\varphi}}{\partial \hat y^3}  \right] = 0 . 
\label{generalisedorrsommerfeldintaylorvariables3}
\end{eqnarray}
\\  
{\bf The generalised Squire equation for large swirl}
\begin{eqnarray}
\frac{\partial^2 \hat u_{\varphi}}{ \partial \hat t \partial \hat x}  + \frac{1}{\sqrt{(1+S^2)}} U_{Gx} \left[Re \frac{\partial^2 \hat u_{\varphi}}{\partial \hat x^2} \right] + \frac{1}{\sqrt{(1+S^2)}} \frac{dU_{Gx}}{dy} \left[ \epsilon_R \frac{\partial \hat u_r}{\partial \hat x} \right]     \nonumber   \\  
 + \frac{S}{\sqrt{(1+S^2)}} U_{G \varphi} \left[ \epsilon_R Re \frac{\partial ^2 \hat u_{\varphi}}{\partial \hat \varphi \partial \hat x} \right]  
 + \left[\frac{\partial^3 \hat u_{\varphi}}{\partial \hat x^3} + \frac{\partial^3 \hat u_{\varphi}}{\partial \hat x \partial \hat y^2} \right] + \epsilon_R \frac{\partial ^2 \hat u_{\varphi}}{\partial \hat x \partial \hat y} = 0. 
\label{generalisedsquireintaylorvariables3}
\end{eqnarray}
{\bf The Continuity equation for large swirl}
\begin{eqnarray}
 \frac{\partial \hat u_r}{\partial \hat y}+ \epsilon_R \hat u_r + Re \frac{\partial \hat u_{\varphi}}{\partial \hat \varphi} + \frac{\partial \hat u_x}{\partial \hat x}  = 0. 
\label{generalisedcontinuityintaylorvariables3}
\end{eqnarray}

It may be verified that, proceeding now to the limit $S \rightarrow \infty$, the above set of equations, 
(\ref{generalisedorrsommerfeldintaylorvariables3}, \ref{generalisedsquireintaylorvariables3}, \ref{generalisedcontinuityintaylorvariables3}) 
get reduced for toroidal disturbances, $\frac{\partial}{\partial \varphi} = 0$, to the equations set up by \cite{taylorscientificpapers} for the classical problem of stability of the 
flow in the gap between rotating cylinders. It may be noted in this context that $Re^2$ may be replaced by the Taylor number $Ta$, and $Re$ by $\sqrt{Ta}$ since $Re^2  = Ta$, see eg. 
\cite{criminalejacksonjoslin}.

\section{Discussion, Conclusions and Outlook}
The set of {\it generalised Orr-Sommerfeld, Squire and Continuity} equations derived, (\ref{generalisedossqcont}) or (\ref{generalisedorrsommerfeld}, \ref{generalisedsquire}, \ref{generalisedcontinuity}) 
with  homogeneous boundary conditions for $\mathbf A$, is an eigenvalue problem for the complex frequency $\omega = \omega_r + \imath \omega_i$ that depends upon the three parameters, $(Re, S, \epsilon_R)$ and the mode of the 
disturbance, $n_{\varphi}$. In the present state of the art this eigenvalue problem is not tractable through analytical means that extract the nature of the dependence of the 
eigenvalues on the parameters, making it inevitable to obtain this information only numerically. The task then remaining is to extract the {\it globally critical Reynolds number} 
out of the {\it neutrally stable surfaces} defined through $\omega _i (Re, S, \epsilon_R) = 0$ over the entire range of possible modes $n_{\varphi}$. The computational task involved 
is too voluminous to be accomodated within the present short paper. It is therefore meaningful to 
focus attention in this paper on such inferences on the characteristics of transition that can be drawn directly from 
the equations themselves. 

A point that seems to have been overlooked in published literature up to now is that in both the limiting cases, which belong to classical problems of flow transition,  
the set of known and well established equations governing small-amplitude disturbance propagation may be understood as approximations of one and the same set of equations. 
These are obtainable from the set of {\it generalised Orr-Sommerfeld, Squire and Continuity} for basic flows with helical streamlines viewed in the revised parameter space. The 
well known equations for the particular cases of transition in plane channel flow and for {\it Taylor instability} in the flow in the gap between concentric circular cylinders 
with rotation of the inner cylinder are both recoverable from the {\it generalised Orr-Sommerfeld, Squire and Continuity equations} on setting the {\it Swirl parameter}, $S$, to 
either $0$ or $\infty$ respectively. 

A feature of interest brought out by the set of {\it generalised Orr-Sommerfeld, Squire and Continuity equations} themselves follows from a cursory inspection of the coefficient 
of the second derivative of $A_r$ in (\ref{reorderedgeneralisedossqcont}), analogous to the case of the basic flow with straight and parallel streamlines. A {\it critical layer} 
may then arise when transition occurs at a sufficiently high Reynolds number and the coefficient of the second derivative of $A_r$ passes through zero. It is worth noting in this 
context that it is the Reynolds number based on $U_{ref}$ that is decisive, and this depends upon both the characteristic velocities of the problem on hand, {\it viz.} $U_{refx}$ 
and $U_{ref \varphi}$. It follows herefrom that the {\it location of the critical layer} is governed not merely by the swirl and geometric parameters, $S$ and $\epsilon_R$, but 
also by the {\it mode of the disturbance}, describable through $n_{\varphi}$. Furthermore, the {\it thickness of the critical layer} scales with $Re^{-\frac{1}{3}}$ a result that, 
although formally is the same as in the classical case of no swirl, expresses dependence on swirl since $Re$ is formed with $U_{ref}$ which in turn depends upon both the 
characteristic velocities  $U_{refx}$ and $U_{ref \varphi}$.

The question that then arises, as to whether the {\it transition initiating disturbance of the spiral flow at given values of $S$ and $\epsilon_R$} is {\it toroidal} or 
{\it helical}, remains open at this stage. Answering this question calls for a comparison of the {\it local critical Reynolds numbers}, which are neutrally stable points in the 
($S, \epsilon_R$)-space at which $\frac{dRe}{d \lambda_x} = 0$, over the range of interest of $S$ and $\epsilon_R$ for different modes of disturbance $n_{\varphi}$. Such a 
comparison of the {\it local critical Reynolds numbers} calls first for obtaining values for the {\it local critical Reynolds numbers} over a wide range of values of 
($S, \epsilon_R$) for different modes, $n_\varphi$, which, in the present state of the art, can be done only numerically.
 

The finding from the present analysis may be summaised as follows: \\ The conclusion that the {\it critical layer}, when it arises, is located at a wall distance where the axial 
flow velocity and the disturbance wave speed equal each other holds only when the transition initiating disturbance is toroidal in nature, not helical. Whether the 
{\it transition initiating disturbance} is toroidal $(n_{\varphi}=0)$ or helical and with which value of $n_{\varphi} \ne 0$, would depend upon the {\it Swirl parameter}, $S$, and 
the ratio of the gap width to the mean radius, $\epsilon_R$, and this requires numerically solving the eigenvalue problem and analysing the solutions obtained.

 The author gratefully acknowledges the benefit of discussions with Professors Roddam Narasimha, FRS, K. R. Sreenivasan and Jeanette Hussong on the subject of this paper.

%
\section*{Appendix: The generalised Orr-Sommerfeld, Squire and Continuity equations in fully developed annular spiral flows}

{\bf The generalised Orr-Sommerfeld equation}
\begin{eqnarray}
\frac{\partial}{\partial t} \left[-\frac{\partial^2 u_r}{\partial r^2} - \frac{\partial^2 u_r}{\partial x^2} - \frac{1}{r} \frac{\partial u_r}{\partial r} + \frac{1}{r} \frac{\partial^2 u_{\varphi}}{\partial r \partial \varphi} - \frac{1}{r} \frac{\partial u_{\varphi}}{\partial \varphi} + O(r^{-2}) \right] \nonumber  \\ 
+ \frac{1}{\sqrt{(1+S^2)}} U_{Gx}\left[\frac{1}{r}\frac{\partial^3 u_r}{\partial r^2 \partial \varphi } - \frac{\partial^3 u_r}{\partial r^2 \partial x} - \frac{1}{r} \frac{\partial^2 u_r}{\partial r \partial x} - \frac{1}{r} \frac{\partial^2 u_{\varphi}}{\partial r \partial \varphi}   - \frac{\partial^3 u_r}{\partial x^3} + O(r^{-2}) \right]  \nonumber  \\ 
+ \frac{1}{\sqrt{(1+S^2)}} \frac{d U_{Gx}}{dr} \left[  \frac{1}{r} \frac{\partial u_r}{\partial x} + \frac{1}{r} \frac{\partial^2 u_{\varphi}}{\partial \varphi \partial x} + O(r^{-2}) \right] + \frac{1}{\sqrt{(1+S^2)}} \frac{d^2 U_{Gx}}{dr^2} \frac{\partial u_r}{\partial x}   \nonumber   \\ 
%
+ \frac{S}{\sqrt{(1+S^2)}} U_{G \varphi} \left[-\frac{1}{r} \frac{\partial ^3 u_r}{\partial r^2 \partial \varphi} - \frac{1}{r} \frac{\partial ^3 u_r}{\partial x^2 \partial \varphi} + \frac{2}{r} \frac{\partial ^2 u_\varphi}{\partial x^2} + O(r^{-2}) \right]   \nonumber   \\ 
+ \frac{S}{\sqrt{(1+S^2)}} \frac{d U_{G \varphi}}{dr} \left[\frac{1}{r} \frac{\partial^2 u_r}{\partial r \partial \varphi} + \frac{1}{r} \frac{\partial^2 u_x}{\partial x \partial \varphi} + O(r^{-2}) \right] + \frac{S}{\sqrt{(1+S^2)}} \frac{d^2 U_{G \varphi}}{d r^2} \left[ \frac{1}{r} \frac{\partial u_r}{\partial \varphi} \right]  \nonumber   \\  
+ \frac{1}{Re} \left[\frac{\partial^4 u_r }{\partial x^4} + \frac{\partial^4 u_r}{\partial r^4} + 2 \frac{\partial^4 u_r}{\partial x^2 \partial r^2} + \frac{2}{r} \frac{\partial^3 u_r}{\partial r \partial x^2} + \frac{2}{r}\frac{\partial^3 u_{\varphi} }{\partial r^3} \right] = 0 . 
\label{generalisedorrsommerfeld}
\end{eqnarray}
{\bf The generalised Squire equation}
\begin{eqnarray}
 \frac{\partial}{\partial t} \left[\frac{\partial u_{\varphi}}{\partial x} - \frac{1}{r} \frac{\partial u_x}{\partial \varphi} \right] + \frac{1}{\sqrt{(1+S^2)}} U_{Gx} \left[\frac{\partial^2 u_{\varphi}}{\partial x^2} + \frac{1}{r} \frac{\partial u_r}{\partial r} + O(r^{-2}) \right] \nonumber \\  
+\frac{1}{\sqrt{(1+S^2)}} \frac{dU_{Gx}}{dr} \left[\frac{\partial u_r}{\partial x} \right]   \nonumber   
 + \frac{S}{\sqrt{(1+S^2)}} U_{G \varphi} \left[ \frac{1}{r} \frac{\partial ^2 u_{\varphi}}{\partial \varphi \partial x} - \frac{1}{r} \frac{\partial ^2 u_x}{\partial \varphi ^2} + \frac{1}{r} \frac{\partial u_r}{\partial x} \right] \nonumber \\ 
- \frac{S}{\sqrt{(1+S^2)}} \frac{d U_{G \varphi}}{dr} \left[\frac{1}{r} \frac{\partial u_r}{\partial \varphi} \right]  \nonumber    
+\frac{1}{Re} \left[\frac{\partial^3 u_{\varphi}}{\partial x^3} + \frac{\partial^3 u_{\varphi}}{\partial  x \partial r^2} + \frac{1}{r}\frac{\partial^2 u_{\varphi}}{\partial x \partial r} - \frac{1}{r} \frac{\partial^3 u_x}{\partial x^2 \partial \varphi}-\frac{1}{r}\frac{\partial^3 u_x}{\partial r^2 \partial \varphi} \right]=0. \nonumber  \\ 
\label{generalisedsquire}
\end{eqnarray}
{\bf The Continuity equation}
\begin{eqnarray}
 \frac{1}{r} \frac{\partial (r u_r)}{\partial r} + \frac{1}{r} \frac{\partial u_{\varphi}}{\partial \varphi} + \frac{\partial u_x}{\partial x} = 0. 
\label{generalisedcontinuity}
\end{eqnarray}


\begin{thebibliography}{99}
%

%
%
%
\bibitem{albumvandyke}
        \textsc{Van Dyke, M.}
        \textit{An Album of Fluid Motion}. Parabolic Press, Stanford University, California, USA, 1982.
%
 
\bibitem{stuart}
        \textsc{Stuart, J.~T.} Hydrodynamic Stability. Chap. IX in \textit{Laminar Boundary Layers (Ed. L. Rosenhead)}, Oxford University Press, 1963 (Dover Ed. 1988)
%

\bibitem{cclin} 
       \textsc{Lin, C.~C.} \textit{The Theory of Hydrodynamic Stability}. Cambridge University Press, Cambridge, UK, 1966.
%
\bibitem{drazinreid}
       \textsc{Drazin, P.~G. \& Reid, W.~H.} \textit{Hydrodynamic stability}. 
       \emph{Cambridge University Press}, 1981.
%
\bibitem{maslowe1} \textsc{Maslowe, S. A.} Shear Flow Instabilities and Transition. In \textit{Hydrodynamic 
 Instabilities and the Transition to Turbulence (Eds.: Swinney, H. L. and Gollub, J. P.)}, Springer 
 Verlag, 1981, Chapt. 7, 181-228.
%
%
%
\bibitem{maslowe2} \textsc{Maslowe,S. A.} Critical layers in shear flows. In Annual Review of Fluid Mechanics, Vol.18, 1986, pp 105-132. 
%
%
\bibitem{schmidhenningson}
       \textsc{Schmid, P.~J. \& Henningson, D.~S.}. 
Stability and Transition in Shear Flows.
\emph{Springer, Berlin, Applied Mathematical Sciences} {\bf 142}, 1990.
%
\bibitem{ramagovindarajan} \textsc{Rama Govindarajan}. The role of the critical layer in the stability of viscous 
shear flow. \emph{Current Science, Special Section: Instabilities, Transition and Turbulence} \textbf{79}, 6, 2000, 741-746.
%
%
\bibitem{criminalejacksonjoslin}
\textsc{Criminale, W.~O., Jackson, T.~L. \& Joslin, R.~D }
\textit{Theory and Computation in Hydrodynamic Stability}. Cambridge University Press, Cambridge, UK, 2003.  
%

%

\bibitem{chandrasekhar}
         \textsc{Chandrasekhar, S.} Hydrodynamic and Hydromagnetic Stability. 
         \emph{Dover}, 1981. 
%
\bibitem{schlichting} 
        \textsc{Schlichting, H.} \textit{Grenzschicht-Theorie}, \emph{Verlag G. Braun}, Karlsruhe, Germany 1951. See also Boundary-Layer Theory. 7th Edition, McGraw-Hill, 1979 and 
        \textsc{Schlichting, H. \& Gersten, K.} Boundary-Layer Theory. 8th Revised and Enlarged Edition, Springer-Verlag, Berlin, 2000
%

 
%
 
%
  
\bibitem{chossatiooss}
        \textsc{Chossat, P. \& Iooss, G.} 
        \textit{The Couette-Taylor Problem}. Springer, Berlin, 1992.
%

%
%
\bibitem{meyerspasche}
        \textsc{Meyer-Spasche, R.} 
        \textit{Pattern Formation in Viscous Flows}. Birkh\"auser Verlag, Basel, 1991.
%
  
%
%
\bibitem{hasoonmartin} 
         \textsc{Hasoon, M.~A. \& Martin B.~W.} The stability of viscous axial flow in an annulus with a rotating inner cylinder. \emph{Proc. Royal Soc. London A} \textbf{352}, 1977, 351-380
%
%
\bibitem{diprimapridor}
         \textsc{DiPrima, R. C. \& Pridor, A.} The stability of viscous flow between rotating concentric cylinders with an axial flow. \emph{Proc. Royal Soc. London A} \textbf{366}, 1979, 555-573
%
\bibitem{takeuchijankowski}
        \textsc{Takeuchi, D.~I. \& Jankowski, D.~F.} A numerical and experimental investigation of the stability of spiral Poiseuille flow. 
        \emph{J. Fluid Mech.}\textbf{102}, 1981, 101-126.
%

%
%
 \bibitem{ngturner}
         \textsc{Ng, B.~S. \& Turner E.~R.} 1982, On the linear stability of spiral flow between rotating cylinders. \emph{Proc. Royal Soc. London A} \textbf{382}, 1982, 83-102
%
 \bibitem{cotrellpearlstein} 
         \textsc{Cotrell, D.~L. \& Pearlstein, A.~J.} The connection between centrifugal instability and Tollmien-Schlichting-like instability for spiral Poiseuille flow, 
         \emph{J. Fluid Mech.}\textbf{509}, 2004, 331-351.  
%

%
%
\bibitem{meseguermarques1}
        \textsc{Meseguer, A. \& Marques, F.} On the competition between centrifugal and shear instability in spiral Couette flow. 
        \emph{J. Fluid Mech.}\textbf{402}, 2000, 33-56.  
%
\bibitem{meseguermarques2} 
        \textsc{Meseguer, A. \& Marques, F.} On the competition between centrifugal and shear instability in spiral Poiseuille flow. \emph{J. Fluid Mech.}\textbf{455}, 2002, 129-148. 
%
\bibitem{meseguermarques3} 
        \textsc{Meseguer, A. \& Marques, F.} On the stability of medium gap corotating spiral Poiseuille flow. \emph{Physics of Fluids} \textbf{17}, 2005, 094104
%
\bibitem{avilameseguermarques} 
        \textsc{Avila, M., Meseguer, A. \& Marques, F.} Double Hopf bifurcation in corotating spiral Poiseuille flow. \emph{Physics of Fluids} \textbf{18}, 2006
%
\bibitem{meseguermellibovskyavilamarques}
        \textsc{Meseguer, A., Mellibovsky, F. Avila, M. \& Marques, F.} 2009 Instability mechanism and transition scenarios of spiral turbulence in Taylor-Couette flow. 
        \emph{Physical Review} \textbf{E 80}, 2009, 046315
%
\bibitem{marquesmeseguerlopezetal} 
        \textsc{Marques, F, Meseguer, A., Lopez, J.~M., Rafael Pacher, J. \& Lopez, J.~A.} Bifurcations with imperfect SO(2) symmetry and pinning of rotating waves. 
        \emph{Proc. R. Soc. London A} \textbf{469}, 2013, 469.2152, (2013):20120348
%
 \bibitem{leclercqpierscott}   
         \textsc{Leclerq, C., Pier, B. \& Scott, J.~F.} Temporal stability of eccentric Taylor-Couette-Poiseuille flow. \emph{J. Fluid Mech.}\textbf{733}, 2013, 68-99.  
%
\bibitem{deguchialtmeyer} 
        \textsc{Deguchi, K. and Altmeyer, S.~A.} 2013 Fully nonlinear mode competitions of nearly bicritical spiral in Taylor vortices in Taylor-Couette flow.  
        \emph{Physical Review} \textbf{E 87}, 2013, 043017 
%
\bibitem{altmeyer} 
        \textsc{Altmeyer, S.~A.} 2014 On secondary instabilities generating footbridges between spiral vortex flow. \emph{Fluid Dynamics Research} \textbf{26}, 2014, 2        
%
%
%
%
%
\bibitem{batchelor}
\textsc{Batchelor, G.~K. } \textit{An Introduction to Fluid Dynamics}. Cambridge University Press, Cambridge, UK, 1967.  
%
\bibitem{taylorscientificpapers}
        \textsc{Taylor, G.~I.} Stability of a viscous fluid contained between two rotating cylinders, in \textit{The Scientific Papers of G. I. Taylor (Ed. G. K. Batchelor), Paper no. 5 in Vol. IV}, Cambridge University Press, 1971, 34.85.
%
  
%


\bibitem{whitham} \textsc{Whitham G. B.} \textit{Linear and Nonlinear Waves}. Wiley, 1971 
%
%
%
%
%
%
%





\end{thebibliography}
\end{document}